\documentclass[10pt,a4paper]{article}
\usepackage{amsmath,amssymb,amsfonts}
\usepackage{graphicx}
\usepackage{hyperref}
\usepackage[margin=0.75in]{geometry}
\usepackage{float}
\usepackage{bm}
\usepackage{booktabs}
\usepackage{tabularx}

\usepackage{setspace}
\usepackage{titlesec}
\titlespacing*{\section}{0pt}{8pt}{4pt}
\titlespacing*{\subsection}{0pt}{6pt}{3pt}
\titlespacing*{\subsubsection}{0pt}{4pt}{2pt}

\AtBeginDocument{
  \setlength{\abovedisplayskip}{4pt}
  \setlength{\belowdisplayskip}{4pt}
  \setlength{\abovedisplayshortskip}{2pt}
  \setlength{\belowdisplayshortskip}{2pt}
}

\let\oldbibliography\thebibliography
\renewcommand{\thebibliography}[1]{%
  \oldbibliography{#1}%
  \setlength{\itemsep}{0pt}%
  \setlength{\parskip}{0pt}%
}

\title{An Idealized Delay-Differential Model of Scuba Diver
Porpoising and Runaway Ascent}

\author{
S. H. S. Herho$^{1,2,3,*}$,
F. A. R. Abdullah$^{3,4}$,
I. P. Anwar$^{3,4}$,\\
F. Khadami$^{3,4}$,
A. P. Handayani$^{1}$,
K. A. Sujatmiko$^{4}$,\\
R. Suwarman$^{6}$,
and D. E. Irawan$^{2}$
}

\date{}

\def\Address{$^{1}$Center for Agrarian Studies, Bandung Institute of
Technology, Bandung, West Java, Indonesia\\
$^{2}$Applied Geology Research Group, Bandung Institute of Technology,
Bandung, West Java, Indonesia\\
$^{3}$Indonesian Subaquatic Sport Association (POSSI), North Jakarta,
DKI Jakarta, Indonesia\\
$^{4}$Applied and Environmental Oceanography Research Group, Bandung
Institute of Technology, Bandung, West Java, Indonesia\\
$^{5}$Spatial Systems and Cadaster Research Group, Bandung Institute
of Technology, Bandung, West Java, Indonesia\\
$^{6}$Atmospheric Science Research Group, Bandung Institute of
Technology, Bandung, West Java, Indonesia}

\def\corrAuthor{Corresponding Author}
\def\corrEmail{sandyherho@itb.ac.id}

\begin{document}
\maketitle

\begin{center}
\small
\Address\\[2pt]
$^{*}$\corrAuthor: \href{mailto:\corrEmail}{\corrEmail}
\end{center}

\begin{abstract}
\noindent
A scuba diver holding constant depth balances on an unstable equilibrium:
the gas carried in the suit and buoyancy compensator compresses with depth, so the
buoyant force falls as the diver sinks and rises as the diver ascends. We represent the
diver as a proportional-derivative controller that regulates this compressible-buoyancy
saddle after a finite reaction delay, and we derive the governing delay differential
equation from the vertical force balance and the isothermal gas law, reducing it to a
damping ratio, two control gains, and a dimensionless delay. The characteristic
spectrum, obtained by pseudospectral collocation of the semigroup generator and checked
against a direct Newton solution of the characteristic equation, locates the Hopf
boundary that separates stable hovering from sustained porpoising; for the baseline
diver the critical reaction delay is 3.36 s and the onset period is 28.8 s. The
bifurcation is supercritical, and because the saturating force is the quadratic
hydrodynamic drag, the limit-cycle amplitude grows in proportion to the delay excess
rather than as its square root. The safe-operating envelope shows that runaway ascent
is triggered by saturation of the compensator, not by loss of linear stability, so a
stable and an unstable diver can share the same escape threshold. As onset is
approached, the lag-one autocorrelation and variance rise while the fitted recovery
rate falls and matches the spectral abscissa, giving an eigenvalue-exact early warning
of the transition.
\end{abstract}


\section{Introduction}

Living systems hold themselves at unstable mechanical equilibria by feedback that is
never instantaneous, and the finite latency of that feedback is a dynamical
ingredient in its own right rather than a technical nuisance. When a controller acts
on information that is delayed relative to the state it corrects, the stabilized
equilibrium can lose stability through a Hopf bifurcation, and the system settles into
a self-sustained oscillation whose frequency is set by the delay rather than by any
oscillatory mode of the underlying plant \cite{Stepan2009,Milton2009}. This mechanism
recurs across otherwise unrelated settings in which a human or animal closes a control
loop around an unstable configuration: the delayed stabilization of upright stance and
of the balanced pole, in which reflex latency converts a controllable inverted
pendulum into a swaying limit cycle \cite{Milton2009,Masani2006}, and the delayed
car-following response of a line of drivers, in which reaction time seeds the
stop-and-go waves of highway traffic \cite{Orosz2010}. The mathematical object common
to these systems is a delay differential equation whose state is an infinite-dimensional
history segment and whose stability is governed by a transcendental characteristic
spectrum, so that the tools of infinite-dimensional dynamical systems and local
bifurcation theory apply directly \cite{Diekmann1995,Guckenheimer1983}.

A submerged diver holding depth is an unrecognized member of this class, and a
consequential one. The gas carried in the breathing system, the wetsuit or drysuit,
and the buoyancy compensator compresses with depth according to Boyle's law, so the
buoyant force weakens as the diver sinks and strengthens as the diver rises. The
neutral hovering state is therefore not a stable balance but an unstable one, a
negative-stiffness equilibrium whose linearization in the vertical coordinate is a
saddle rather than a center, in direct analogy with the inverted pendulum of the
balance literature. A diver maintains depth only by continually sensing the depth
error and its rate and adjusting the compensator gas, and that correction is issued
after a finite perceptual and motor delay. When the delay or the control gain grows
too large, depth-keeping degrades into a bounded oscillation that divers call
porpoising, and in the extreme the compensator saturates and the diver departs the
neutral band entirely in an uncontrolled ascent or descent. The safety stakes of this
failure are severe and well documented: emergency or uncontrolled ascent is the single
most common disabling agent in recreational open-circuit diving fatalities and is
associated with the large majority of arterial gas embolism cases, and loss of
buoyancy control is itself a frequent contributing agent \cite{Denoble2008}, while the
rapid pressure reduction of an uncontrolled ascent is the proximate cause of arterial
gas embolism and decompression sickness \cite{Vann2011}. The dynamical event that
precedes these injuries, the loss of stable depth regulation, has nonetheless not been
analyzed as such.

Where buoyancy control has been modeled, it has been treated almost exclusively as an
engineering autopilot problem, in which a fast automatic controller is designed to
regulate the buoyancy of a device. Dynamic models of diver buoyancy compensators, of
variable-buoyancy systems for autonomous underwater vehicles, and of submarine
hovering through the blowing and venting of ballast all pose the question of how to
build a controller that holds depth well \cite{Valenko2016,Font2013,Fossen2011}. That
framing suppresses precisely the feature that is essential here, because a well-designed
machine controller minimizes its own latency, whereas the human diver cannot, and it is
the irreducible human reaction delay acting on an inherently unstable plant that
generates porpoising. To our knowledge, porpoising has not been formulated as a
delay-induced Hopf bifurcation, nor examined with the apparatus now standard for such
problems, namely spectral analysis of the delay differential equation, continuation of
the emergent limit cycle, characterization of the global basin under actuator
saturation, and the detection of onset through critical slowing down and
information-theoretic structure \cite{Scheffer2009,Dakos2012,BandtPompe2002,Rosso2007}.

This study develops that analysis for an idealized model of a scuba diver as a delayed
proportional-derivative regulator of a compressible-buoyancy equilibrium. We derive the
governing delay differential equation from the vertical force balance and the
thermodynamics of the carried gas, reduce it to four dimensionless groups, and obtain
its characteristic spectrum by pseudospectral collocation of the semigroup generator,
which locates the Hopf boundary and the onset frequency to high accuracy
\cite{Breda2005}. We show that the onset of porpoising is a supercritical Hopf
bifurcation whose amplitude is fixed by the quadratic hydrodynamic drag and grows
linearly, rather than as a square root, in the distance to onset, a scaling that
reflects the degree-two homogeneity of the saturating dissipation. We map the global
safe-operating envelope and find that the boundary between recovery and escape is
governed by compensator saturation rather than by the sign of the leading eigenvalue,
so that local stability and global safety are distinct properties. We demonstrate that
the approach to onset is heralded by critical-slowing-down precursors whose fitted
recovery rate coincides with the pseudospectral spectral abscissa, and we place the
hovering, porpoising, and runaway regimes on the complexity-entropy plane. The model is
deliberately minimal, a toy model in the sense that it retains a single mechanism and
discards everything not essential to it, in keeping with a broader program of idealized,
reproducible models of geophysical and physical systems that pair a focused governing
equation with information-theoretic diagnostics \cite{Herho2025KH,Irawan2026KdV}. Its
purpose is not to reproduce the full behavior of a real dive but to explain, in the
language of nonlinear dynamics, why delayed regulation of an inherently unstable
buoyancy equilibrium should give rise to hovering, porpoising, and runaway at all, and
what governs the transitions among them.

\section{Methods}
\subsection{Model Description}

\begin{figure}[H]
\centering
\includegraphics[width=0.62\textwidth]{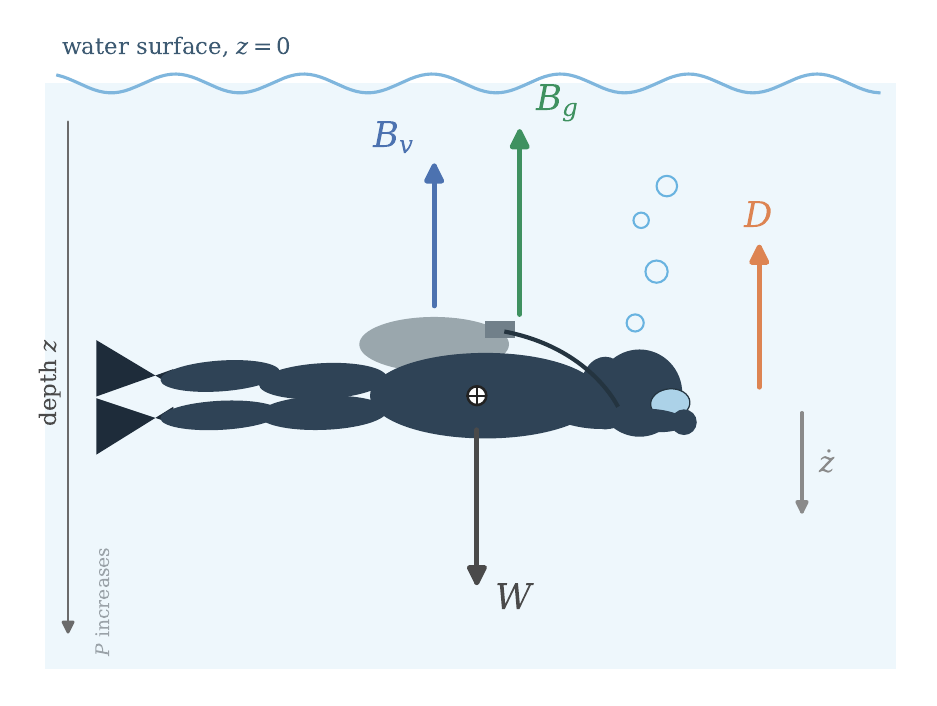}
\caption{Free-body diagram of the diver in vertical motion. Depth $z$ increases
downward along $\hat{\mathbf z}$, and the ambient pressure $P$ increases with it.
The weight $W=mg$ acts downward. The incompressible buoyancy
$B_v=\rho g V_{\rm inc}$, set by the displaced volume of the body and rigid gear,
does not vary with depth. The compressible buoyancy
$B_g=\rho g\,G_{\rm surf}P_0/P(z)$, set by the carried gas, decreases as the diver
descends and provides the destabilizing element. The drag $D$ opposes the vertical
velocity $\dot z$ and reverses sense between descent and ascent.}
\label{fig:fbd}
\end{figure}

The diver is treated as a rigid body in vertical (heave) motion, and only the
vertical degree of freedom is retained. Let $\hat{\mathbf z}$ be the unit vector
directed vertically downward, let $z(t)$ be the depth of the diver's center of
mass measured positive along $\hat{\mathbf z}$ from the free surface, and let
$\dot z$ and $\ddot z$ be its velocity and acceleration. Lateral translation and
body rotation are suppressed on the assumption that the diver holds attitude and
horizontal position, and the attitude control that supports this reduction is not
modeled here. The vertical force balance is shown in Figure~\ref{fig:fbd}.

The ambient pressure follows from hydrostatic balance. In the surrounding fluid at
rest the Cauchy momentum equation reduces to $\nabla P = \rho\,\mathbf g$, where
$P(\mathbf x)$ is the pressure, $\rho$ the seawater density taken uniform, and
$\mathbf g = g\,\hat{\mathbf z}$ the gravitational acceleration of magnitude $g$.
Projecting on $\hat{\mathbf z}$ gives $\partial P/\partial z = \rho g$, and
integration from the free surface, where $P=P_0$ is the surface pressure, yields
\begin{equation}
P(z) = P_0 + \rho g\,z .
\label{eq:pressure}
\end{equation}
Uniform $\rho$ is accurate to the order of the seawater compressibility over the
depth range considered and is used throughout.

The buoyant force is the resultant of the pressure on the wetted surface. Let the
diver occupy a closed region $\mathcal V$ with piecewise-smooth boundary
$\partial\mathcal V$ and outward unit normal $\mathbf n$. Setting aside for the
moment the pressure disturbance produced by the body's own motion, the fluid
exerts on the body the force
\begin{equation}
\mathbf F_{P} = \oint_{\partial\mathcal V}\bigl(-P\,\mathbf n\bigr)\,dS
             = -\int_{\mathcal V}\nabla P\,dV ,
\label{eq:gradient_theorem}
\end{equation}
where the second equality is the gradient theorem. With the hydrostatic gradient
$\nabla P = \rho g\,\hat{\mathbf z}$, uniform on the scale of the body,
$\mathbf F_{P} = -\rho g\,\lvert\mathcal V\rvert\,\hat{\mathbf z}$, an upward force
of magnitude $\rho g\,\lvert\mathcal V\rvert$, with $\lvert\mathcal V\rvert$ the
displaced volume. The displaced volume splits into a rigid, depth-invariant part
$V_{\rm inc}$ from the body tissue and hard equipment and a compliant gaseous part
$V_{\rm gas}(z)$ from the wetsuit or drysuit gas together with the buoyancy
compensator. Because the pressure gradient is uniform across the envelope, the two
buoyancies superpose,
\begin{equation}
B = B_v + B_g = \rho g\,V_{\rm inc} + \rho g\,V_{\rm gas}(z) .
\label{eq:buoyancy_split}
\end{equation}

The compressible part follows from the thermodynamics of the sealed gas, modeled as
a polytropic process $P\,V_{\rm gas}^{\,n} = \text{const}$, with polytropic index
$n$. The bounding cases are isothermal compression, $n=1$, in which the gas holds
the temperature of the surrounding water, and adiabatic compression, $n=c_p/c_v$,
in which it exchanges no heat. Buoyancy adjustment during depth-keeping proceeds
over seconds to tens of seconds, which is long relative to the thermal relaxation
time of the thin gas layers in the suit and compensator against the large water
reservoir in contact with them, so the compression is isothermal to good
approximation and we set $n=1$. The gas is then referenced to its volume at the
surface, where the pressure is $P_0$, and writing $G_{\rm surf}$ for that
surface-referenced volume,
\begin{equation}
V_{\rm gas}(z) = G_{\rm surf}\,\frac{P_0}{P(z)},
\qquad
B_g = \rho g\,G_{\rm surf}\,\frac{P_0}{P(z)} .
\label{eq:boyle}
\end{equation}
Retaining a general index would multiply $V_{\rm gas}$ by $(P_0/P)^{1/n-1}$ and,
through the linearization below, scale the buoyancy stiffness by $1/n$; the
isothermal closure fixes that factor at unity, and \eqref{eq:boyle} is the state
relation used in the solver. The surface-referenced load partitions into the suit
gas $V_{s0}$, held fixed during a dive, and the compensator gas $q$, which the
diver regulates, so that $G_{\rm surf} = V_{s0} + q$.

The reaction of the fluid to the body's motion separates into an inertial part and
a dissipative part. In the incompressible potential-flow approximation the
disturbance field is $\mathbf u = \nabla\Phi$ with
$\Phi(\mathbf x,t) = \dot z(t)\,\varphi(\mathbf x)$, where the unit potential
$\varphi$ solves $\nabla^2\varphi = 0$ in the fluid domain $\Omega_f$ subject to
$\partial_n\varphi = \hat{\mathbf z}\cdot\mathbf n$ on $\partial\mathcal V$ and
$\nabla\varphi\to\mathbf 0$ far from the body. The kinetic energy of the entrained
fluid is quadratic in the body speed,
\begin{equation}
T_f = \tfrac12\rho\int_{\Omega_f}\lvert\nabla\Phi\rvert^{2}\,dV
    = \tfrac12\Bigl(\rho\!\int_{\Omega_f}\lvert\nabla\varphi\rvert^{2}\,dV\Bigr)\dot z^{2}
    \equiv \tfrac12\,m_a\,\dot z^{2},
\label{eq:fluid_ke}
\end{equation}
which defines the added mass $m_a>0$. Taking $m_a$ constant over the depth range
considered, the inertial reaction on the body along $\hat{\mathbf z}$ is
$-\tfrac{d}{dt}\bigl(\partial T_f/\partial\dot z\bigr) = -\,m_a\ddot z$, so the
combined inertia of the body and the entrained fluid is
$m_{\rm eff} = m + m_a$, with $m$ the mass of the diver and equipment. The
dissipative reaction is represented by a linear term plus a quadratic term,
\begin{equation}
D = c_1\,\dot z + \tfrac12\rho\,C_d\,S\,\lvert\dot z\rvert\,\dot z,
\label{eq:drag}
\end{equation}
with $c_1$ a linear drag coefficient for the low-speed and skin-friction regime,
$C_d$ a form-drag coefficient, and $S$ a reference frontal area; the quadratic
term dominates once the Reynolds number $Re = \lvert\dot z\rvert L/\nu$, formed
with a body length $L$ and the kinematic viscosity $\nu$, is appreciable. This
decomposition is standard for submerged bluff bodies \cite{Fossen2011}.

Newton's second law along $\hat{\mathbf z}$, downward positive, reads
$m_{\rm eff}\ddot z = W - B_v - B_g - D$, with $W = mg$ the weight. Introducing the
net weight excess $\Delta \equiv W - B_v = mg - \rho g V_{\rm inc}$, which collects
the two depth-independent contributions, and substituting \eqref{eq:boyle} and
\eqref{eq:drag} gives the nonlinear governing equation of the plant,
\begin{equation}
m_{\rm eff}\,\ddot z
= \Delta - \rho g\,G_{\rm surf}\,\frac{P_0}{P(z)}
  - c_1\,\dot z - \tfrac12\rho\,C_d\,S\,\lvert\dot z\rvert\,\dot z .
\label{eq:eom}
\end{equation}
Equation \eqref{eq:eom} is second order in $z$ once $G_{\rm surf}(t)$ is prescribed
by the control, and it is the continuous model that the numerical treatment
integrates.

Regulation of $G_{\rm surf}$ closes the loop. A human operator senses the depth
error and its rate and actuates the compensator after a finite sensorimotor
latency, and the stabilization of an unstable mechanical equilibrium by delayed
proportional and derivative feedback is well documented for human balance
\cite{Stepan2009,Milton2009,Masani2006}. With $\tau$ the reaction delay, $z^\ast$
the target neutral depth, and $k_p$ and $k_d$ the proportional and derivative
gains, the commanded compensator gas is
\begin{equation}
q_c(t) = q^\ast + k_p\bigl(z(t-\tau) - z^\ast\bigr) + k_d\,\dot z(t-\tau),
\label{eq:control}
\end{equation}
with $q^\ast$ the neutral compensator load fixed below. The compensator volume is
bounded, so the realized load and the resulting surface-referenced gas are
\begin{equation}
q(t) = \operatorname{sat}_{[0,\,q_{\max}]}\!\bigl(q_c(t)\bigr),
\qquad
G_{\rm surf}(t) = V_{s0} + q(t),
\label{eq:saturation}
\end{equation}
where $q_{\max}$ is the compensator capacity at surface pressure and
$\operatorname{sat}_{[a,b]}(\cdot)$ clamps its argument to $[a,b]$. The saturation
in \eqref{eq:saturation} is the amplitude-limiting nonlinearity of the closed loop
and is kept in the model.

A neutral state is a rest state at the target depth. Setting
$\ddot z = \dot z = 0$ and $z = z^\ast$ in \eqref{eq:eom} with the gas at its
neutral load $G^\ast = V_{s0} + q^\ast$ defines
$\mathcal F(z^\ast,G^\ast) = \Delta - \rho g\,G^\ast P_0/P(z^\ast) = 0$. Because
$\partial\mathcal F/\partial G^\ast = -\rho g\,P_0/P(z^\ast)\neq 0$, the
implicit-function theorem gives a unique neutral load for each target depth, and
writing $P_z \equiv P(z^\ast)$,
\begin{equation}
G^\ast = \frac{\Delta\,P_z}{\rho g\,P_0},
\qquad
q^\ast = G^\ast - V_{s0},
\label{eq:equilibrium}
\end{equation}
which is admissible when $\Delta>0$ and $0\le q^\ast\le q_{\max}$.

The behavior near the neutral state follows by linearizing \eqref{eq:eom} under the
delayed control \eqref{eq:control}. Set $z = z^\ast + \xi$, with the depth
perturbation $\xi$ and its derivatives small, and expand the compressible buoyancy
using \eqref{eq:pressure},
\begin{equation}
\frac{P_0}{P(z)} = \frac{P_0}{P_z}\,\frac{1}{1 + \rho g\,\xi/P_z}
= \frac{P_0}{P_z}\left[\,1 - \frac{\rho g}{P_z}\,\xi
+ \Bigl(\frac{\rho g}{P_z}\Bigr)^{2}\xi^{2} - \cdots\right],
\label{eq:pressure_expansion}
\end{equation}
convergent for $\lvert\rho g\,\xi/P_z\rvert < 1$, a condition met for excursions
small against the absolute-pressure scale height $P_z/\rho g$. Holding the gas at
$G^\ast$, the term linear in $\xi$ defines the buoyancy stiffness
\begin{equation}
\beta \equiv -\left.\frac{\partial B_g}{\partial z}\right|_{z^\ast,G^\ast}
      = \frac{(\rho g)^{2}\,G^\ast\,P_0}{P_z^{2}} \;>\; 0 .
\label{eq:beta}
\end{equation}
A downward displacement lowers the compressible buoyancy, so the net downward force
gains a component $+\beta\,\xi$ that reinforces the displacement, and $\beta$ acts
as a negative stiffness that renders the neutral state statically unstable without
control. The sensitivity of the buoyancy to the regulated gas is the control
authority
\begin{equation}
\gamma \equiv \left.\frac{\partial B_g}{\partial G_{\rm surf}}\right|_{z^\ast}
      = \frac{\rho g\,P_0}{P_z},
\label{eq:gamma}
\end{equation}
so the control enters as
$\delta G_{\rm surf} = k_p\,\xi(t-\tau) + k_d\,\dot\xi(t-\tau)$ from
\eqref{eq:control}. The quadratic drag contributes nothing at linear order about
$\dot z = 0$, since $\tfrac{d}{d\dot z}\bigl(\lvert\dot z\rvert\dot z\bigr)
= 2\lvert\dot z\rvert$ vanishes there, leaving the linear drag $c_1\dot\xi$.
Collecting these terms in \eqref{eq:eom} gives the linearized closed loop as the
delay differential equation
\begin{equation}
m_{\rm eff}\,\ddot\xi + c_1\,\dot\xi - \beta\,\xi
+ \gamma k_p\,\xi(t-\tau) + \gamma k_d\,\dot\xi(t-\tau) = 0 .
\label{eq:linear_dde}
\end{equation}
The terms dropped in \eqref{eq:linear_dde} are the quadratic contribution from
\eqref{eq:pressure_expansion} and the quadratic drag from \eqref{eq:drag}, which
are the nonlinearities that bound the amplitude of a self-sustained oscillation and
are kept in the full model \eqref{eq:eom}.

The natural rate of the instability is
\begin{equation}
\omega_0 = \sqrt{\beta / m_{\rm eff}} ,
\label{eq:omega0}
\end{equation}
and rescaling time by it through $s = \omega_0 t$, with the perturbation written in
units of an arbitrary reference length $\ell$ as $\xi = \ell\,x(s)$, reduces
\eqref{eq:linear_dde} to a form set by four dimensionless groups. With
$\dot\xi = \ell\omega_0 x'$ and $\ddot\xi = \ell\omega_0^{2} x''$, a prime denoting
$d/ds$, and dividing by $m_{\rm eff}\ell\omega_0^{2} = \beta\ell$, the groups are
\begin{equation}
\zeta = \frac{c_1}{2\,m_{\rm eff}\,\omega_0}, \qquad
\kappa_p = \frac{\gamma\,k_p}{\beta}, \qquad
\kappa_d = \frac{\gamma\,k_d\,\omega_0}{\beta}, \qquad
\theta = \omega_0\,\tau ,
\label{eq:nondim_groups}
\end{equation}
in which $\zeta$ is a damping ratio, $\kappa_p$ and $\kappa_d$ are the
nondimensional proportional and derivative control authorities, and $\theta$ is the
delay in units of the instability time scale. The linearized closed loop then reads
\begin{equation}
x'' + 2\zeta\,x' - x + \kappa_p\,x(s-\theta) + \kappa_d\,x'(s-\theta) = 0 ,
\label{eq:nondim_dde}
\end{equation}
where the negative coefficient of $x$ carries the destabilizing stiffness
\eqref{eq:beta} and the two retarded terms carry the diver's regulating action.

Equation \eqref{eq:nondim_dde} is a retarded functional differential equation whose
state at time $s$ is the history segment $x_s(\vartheta) = x(s+\vartheta)$ for
$\vartheta\in[-\theta,0]$, an element of the Banach space
$C\bigl([-\theta,0],\mathbb R^{2}\bigr)$ of continuous $(x,x')$ histories. The
modal substitution $x(s) = e^{\mu s}$ produces the characteristic function
\begin{equation}
\chi(\mu) \equiv \mu^{2} + 2\zeta\,\mu - 1
+ \bigl(\kappa_p + \kappa_d\,\mu\bigr)e^{-\mu\theta} = 0 ,
\label{eq:characteristic}
\end{equation}
a transcendental quasi-polynomial with countably many roots $\{\mu_k\}$, the
spectrum of the generator of the solution semigroup. Asymptotic stability of the
neutral state is governed by the rightmost root, and stability is lost when a
complex conjugate pair crosses the imaginary axis, which is the mechanism that
produces the bounded oscillation of interest. In dimensional variables
\eqref{eq:characteristic} corresponds to
$m_{\rm eff}\lambda^{2} + c_1\lambda - \beta
+ (\gamma k_p + \gamma k_d\lambda)e^{-\lambda\tau} = 0$ with $\lambda = \omega_0\mu$.

The uncontrolled limit shows why active, delayed regulation is required. Setting
$\kappa_p = \kappa_d = 0$ in \eqref{eq:nondim_dde}, which holds the gas fixed at its
neutral load, leaves
\begin{equation}
x'' + 2\zeta\,x' - x = 0 ,
\label{eq:openloop}
\end{equation}
a linear oscillator with negative stiffness whose characteristic roots are
$\mu_\pm = -\zeta \pm \sqrt{\zeta^{2}+1}$. Since $\sqrt{\zeta^{2}+1} > \zeta$ for
every $\zeta \ge 0$, the root $\mu_+$ is positive and $\mu_-$ negative, so the
neutral depth is a saddle of the uncontrolled plant and a small perturbation
diverges along the unstable manifold at the rate $\omega_0\,\mu_+$. The governing
model of the diver is therefore the nonlinear plant \eqref{eq:eom}, closed by the
saturated delayed regulation \eqref{eq:control}--\eqref{eq:saturation}, with
linearization \eqref{eq:nondim_dde} and characteristic function
\eqref{eq:characteristic}. This is the continuous formulation on which the analysis
is built.

\subsection{Numerical Implementation}

The analysis proceeds entirely from the governing model \eqref{eq:eom} and its
linearization \eqref{eq:nondim_dde}, without recourse to an external solver or to
stored trajectory data. Two numerical treatments carry the work. The first returns
the characteristic spectrum of the linearized delay equation by discretizing the
generator of its solution semigroup, and the second advances the nonlinear plant in
time on a mesh aligned to the delay. The design follows a program of open-source,
just-in-time-accelerated solvers for idealized geophysical and physical models in
which a focused governing equation is paired with a scheme matched to its regime and
with reproducible, self-describing output \cite{Herho2025KH,Herho2026Wave,Irawan2026KdV}.
The computations are performed in double precision using NumPy for array operations
\cite{Harris2020}, SciPy for root finding and rank statistics \cite{Virtanen2020},
and Matplotlib for rendering \cite{Hunter2007}, with the two performance-critical
kernels, the spectral matrix assembly and the time-stepping loop, compiled to machine
code by Numba \cite{Lam2015}; theoretical reference curves for the ordinal diagnostics
are drawn from \texttt{ordpy} \cite{Pessa2021}.

The stability of the neutral state is governed by the spectrum of the linear
operator associated with \eqref{eq:nondim_dde}. Written as a first-order system with
$y=(x,x')^{\top}$, the linearization reads $y'(s)=L_0\,y(s)+L_1\,y(s-\theta)$, where
\begin{equation}
L_0 = \begin{pmatrix} 0 & 1 \\ 1 & -2\zeta \end{pmatrix}, \qquad
L_1 = \begin{pmatrix} 0 & 0 \\ -\kappa_p & -\kappa_d \end{pmatrix}.
\label{eq:state_space}
\end{equation}
The solution over one delay is generated by the strongly continuous semigroup
$\{T(s)\}_{s\ge 0}$ acting on the state space
$X=C\bigl([-\theta,0],\mathbb{C}^{2}\bigr)$ of continuous history segments
$y_s(\vartheta)=y(s+\vartheta)$, and its infinitesimal generator $\mathcal A$ is the
derivative operator
\begin{equation}
(\mathcal A\psi)(\vartheta) = \psi'(\vartheta), \quad \vartheta\in[-\theta,0),
\qquad
(\mathcal A\psi)(0) = L_0\,\psi(0) + L_1\,\psi(-\theta),
\label{eq:generator}
\end{equation}
defined on the domain of continuously differentiable histories that satisfy the
boundary condition in \eqref{eq:generator}. The eigenvalues of $\mathcal A$ are
exactly the roots $\mu$ of the characteristic function \eqref{eq:characteristic},
and asymptotic stability of the neutral state is equivalent to the spectral
abscissa $\alpha=\sup_k\operatorname{Re}\mu_k$ being negative. The operator is
approximated on the Chebyshev-Gauss-Lobatto nodes
$\vartheta_j=\tfrac{\theta}{2}\bigl(\cos(\pi j/N)-1\bigr)$ for $j=0,\dots,N$, which
map the interval $[-\theta,0]$ so that $\vartheta_0=0$ carries the boundary
condition and $\vartheta_N=-\theta$ the fully retarded state. The pseudospectral
collocation of a differential operator on Chebyshev nodes parallels the Fourier
pseudospectral treatment used for dispersive wave dynamics in a companion solver
\cite{Irawan2026KdV}. Writing $D\in\mathbb{R}^{(N+1)\times(N+1)}$ for the barycentric
spectral differentiation matrix on these nodes \cite{Trefethen2000} and rescaling it
to the mapped interval as $D_\theta=(2/\theta)D$, the generator is represented by
$M=D_\theta\otimes I_2$ with its first block row replaced by the boundary condition,
\begin{equation}
M_{[0,:]} = 0,\qquad
M_{[0,0]} = L_0,\qquad
M_{[0,N]} = L_1,
\label{eq:collocation}
\end{equation}
so that the discrete eigenproblem $M\,\hat y=\mu\,\hat y$ approximates the spectrum
of $\mathcal A$. The pseudospectral construction inherits the exponential
convergence of Chebyshev collocation for the smooth eigenfunctions of interest
\cite{Breda2005}, and the eigenvalues obtained from \eqref{eq:collocation}, ordered
by decreasing real part, converge geometrically in $N$. Taking $N$ in the range
$24$ to $30$ places the rightmost roots at the level of the working precision, and
these roots agree with a Newton iteration applied directly to
\eqref{eq:characteristic} to a residual of order $10^{-13}$. From this spectrum we
obtain the spectral abscissa as a function of the dimensionless groups, the Hopf
locus as its zero set, and, at a boundary point where a conjugate pair
$\mu=\pm i\Omega$ crosses the imaginary axis, the onset frequency $\Omega$ and the
associated period $2\pi/(\omega_0\Omega)$ through the natural time scale
\eqref{eq:omega0}. The critical delay at which stability is lost is the root of the
spectral abscissa, isolated by the Brent bracketing method \cite{Virtanen2020}, and
the closed-form Hopf relation obtained by separating \eqref{eq:characteristic} at
$\mu=i\Omega$ into its real and imaginary parts,
\begin{equation}
\kappa_p\cos\Omega\theta + \kappa_d\,\Omega\sin\Omega\theta = \Omega^{2}+1,
\qquad
-\kappa_p\sin\Omega\theta + \kappa_d\,\Omega\cos\Omega\theta = -2\zeta\Omega,
\label{eq:hopf_conditions}
\end{equation}
furnishes an independent parametrization of the boundary against which the
collocation result is checked.

The nonlinear plant \eqref{eq:eom} is integrated by a continuous method of steps
\cite{BellenZennaro2003}. The mesh spacing is fixed at $\Delta t=\tau/m$ for an
integer $m$, so the retarded argument evaluated at any whole Runge-Kutta stage
coincides with a stored node and the primary discontinuity points at
$t=\tau,2\tau,\dots$, where successive derivatives of the solution fail to match
across the delay, fall exactly on mesh nodes. Denoting the state
$w=(z,\dot z)^{\top}$ and the right-hand side of \eqref{eq:eom} by
$f\bigl(t,w(t),w(t-\tau)\bigr)$, the classical fourth-order Runge-Kutta formula
advances $w_i\mapsto w_{i+1}$ over $[t_i,t_i+\Delta t]$, and the retarded state
required by the two midpoint stages at $t_i+\Delta t/2$ is supplied by the cubic
Hermite continuous extension of the stored solution across the bracketing retarded
nodes $t_{i-m}$ and $t_{i-m+1}$,
\begin{equation}
\tilde w\bigl(t_i+\tfrac{\Delta t}{2}-\tau\bigr)
= \tfrac12\bigl(w_{i-m}+w_{i-m+1}\bigr)
+ \tfrac{\Delta t}{8}\bigl(f_{i-m}-f_{i-m+1}\bigr),
\label{eq:hermite}
\end{equation}
in which the stored slopes $f_{i-m}$ and $f_{i-m+1}$ are the previously evaluated
right-hand sides at those nodes. Interpolating the retarded argument in this way
keeps it consistent with a genuine continuous extension of the numerical trajectory
rather than a piecewise-constant lookup. The history on $[-\tau,0]$ is the constant
neutral state, and the saturation \eqref{eq:saturation} of the regulated gas is
retained throughout, so that the integrated dynamics include the amplitude-limiting
nonlinearity of the closed loop. Because that constant history renders the solution
only Lipschitz continuous at the origin, the discontinuity propagates across
successive delay intervals and limits the global order of the scheme to second order
in $\Delta t$, below the formal fourth order of the underlying formula, in agreement
with the theory of continuous Runge-Kutta methods for retarded equations
\cite{BellenZennaro2003}. A refinement study in which the mesh is successively
halved confirms this order and establishes that the invariant quantities extracted
from the integrations are independent of the mesh to better than one part in
$10^{5}$, and that the decay rates recovered from transient envelopes reproduce the
pseudospectral abscissa to within a few parts in $10^{3}$. The verification-first
posture adopted here, in which a solver restricted to a regime with an independent
reference is checked end-to-end against that reference before any application, mirrors
the analytical-benchmark validation of related finite-volume and spectral solvers in
this program of work \cite{Herho2025KH,Irawan2026KdV}.

\subsection{Data Analysis}

The quantities derived from the model characterize the neutral state in four
complementary ways: through the spectrum of the linearization, through the amplitude
and the basin of the nonlinear oscillation past onset, through the ordinal structure
of the depth records, and through the signature of critical slowing down as the
onset is approached. This use of information-theoretic descriptors alongside the
classical dynamical quantities continues a diagnostic strategy applied to idealized
flows across this program of work, in which entropy-based and complexity measures
extract organizational structure that first-moment order parameters do not resolve
\cite{Herho2025KH,Irawan2026KdV}. All four groups are computed with NumPy and SciPy
\cite{Harris2020,Virtanen2020} and share the assumptions of the model, namely
constant added mass in \eqref{eq:fluid_ke}, isothermal compression of the carried gas
in \eqref{eq:boyle}, and the saturation \eqref{eq:saturation} of the regulated gas.

The linear characterization is read directly from the rightmost characteristic
root. Its real part, the spectral abscissa, measures the exponential rate of
approach to or departure from neutral and changes sign at the Hopf boundary, and its
imaginary part fixes the frequency of the marginal oscillation and hence the period
of the incipient porpoising through \eqref{eq:omega0}. The distance of the operating
point to the boundary, expressed as the magnitude of the spectral abscissa, provides
a scalar measure of stability margin that orders the diver configurations and, along
a ramp in the reaction delay, marks the crossing into instability.

The nonlinear oscillation is quantified from integrations of the saturated plant.
For a reaction delay beyond onset the depth signal settles onto a bounded limit
cycle, whose steady peak-to-peak amplitude $A$ is measured over the terminal
segment of a long integration once the transient has decayed. The character of the
bifurcation is established by the scaling of this amplitude near onset. Expanding the
normal form of a Hopf bifurcation limited at cubic order gives
$A^{2}\propto(\tau-\tau_c)$ to leading order, so a least-squares regression of the
squared amplitude on the delay excess $\tau-\tau_c$ in a neighborhood of onset,
carried out with \texttt{numpy.polyfit} \cite{Harris2020}, tests supercriticality
through the sign of its slope, with the quadratic drag of \eqref{eq:eom} supplying
the saturating nonlinearity that sets the finite amplitude. The extent of the
recoverable set is mapped separately by advancing an ensemble of initial conditions
spanning the plane of initial depth error and initial vertical velocity and
classifying each start as bounded, when the trajectory returns to a fixed band about
neutral, or as escaping, when the regulated gas saturates and the excursion crosses a
prescribed shallow or deep threshold. The measure of the bounded set and the largest
recoverable depth error at zero initial velocity summarize the safe operating region
for each configuration.

The ordinal characterization assigns to each depth record a pair of
information-theoretic coordinates. Given an embedding dimension $d$ and a lag
$\ell$, each window $\bigl(x_i,x_{i+\ell},\dots,x_{i+(d-1)\ell}\bigr)$ is replaced by
the permutation $\pi$ that sorts it, and the relative frequencies of the $d!$
admissible permutations form the ordinal distribution $P=\{p_\pi\}$
\cite{BandtPompe2002}. The normalized permutation entropy is the Shannon entropy of
this distribution referred to its maximum,
\begin{equation}
H(P) = \frac{1}{\ln d!}\Bigl(-\sum_{\pi} p_\pi \ln p_\pi\Bigr),
\label{eq:perm_entropy}
\end{equation}
and the statistical complexity is the product of this entropy with the
Jensen-Shannon disequilibrium of $P$ from the uniform distribution $U$
\cite{LopezRuiz1995,Rosso2007},
\begin{equation}
C(P) = \frac{\mathcal J(P,U)}{\mathcal J_{\max}}\,H(P),
\qquad
\mathcal J(P,U) = S\!\Bigl(\tfrac{P+U}{2}\Bigr) - \tfrac12 S(P) - \tfrac12 S(U),
\label{eq:complexity}
\end{equation}
with $S(\cdot)$ the Shannon entropy in nats and $\mathcal J_{\max}$ the maximal
Jensen-Shannon divergence, attained when $P$ concentrates on a single pattern. We use
$d=6$ and $\ell=1$, and the estimators are evaluated after sampling each record at a
fixed interval, discarding the initial transient, and adding a small observational
noise to the depth so that the coordinates are formed under realistic conditions. A
white-noise reference occupies the high-entropy, low-complexity corner and a clean
oscillation the low-entropy edge, and the admissible region is bounded by the
extremal complexity curves for the chosen embedding \cite{Pessa2021}; the trajectory
of the coordinate along a ramp in the reaction delay records the migration from the
stochastic corner toward the periodic edge as the neutral state destabilizes.

The approach to onset is detected through critical slowing down. As the reaction
delay tends to its critical value the rightmost root approaches the imaginary axis,
so the relaxation of a perturbation slows, and under weak stochastic forcing the
stationary depth signal acquires higher lag-one autocorrelation and higher variance
while its measured recovery rate falls \cite{Scheffer2009,Dakos2012}. The
autocorrelation and variance are computed from stationary records at a sequence of
delays approaching the boundary, and the recovery rate is estimated by fitting the
logarithm of a deterministic decay envelope over its later portion, where the
rightmost mode dominates, so that the fitted rate approximates $-\omega_0\alpha$ with
$\alpha$ the spectral abscissa. The monotonicity of each indicator with decreasing
distance to the boundary is quantified by the Kendall rank correlation
\cite{Virtanen2020}, a positive value indicating that the indicator rises as onset
is approached, which ties the statistical precursors to the exact spectral distance
provided by the linearization.

Two further analyses assess the sensitivity of these measures to the physical
parameters and their robustness to imperfect observation. The parameter sensitivity
is obtained by perturbing each physical parameter $p$ of the baseline configuration
by a small relative amount and forming the central-difference semi-elasticities
\begin{equation}
\frac{\partial \alpha}{\partial \ln p}
\approx \frac{\alpha\!\left((1+\epsilon)p\right)-\alpha\!\left((1-\epsilon)p\right)}{2\epsilon},
\qquad
\frac{\partial \tau_c}{\partial \ln p}
\approx \frac{\tau_c\!\left((1+\epsilon)p\right)-\tau_c\!\left((1-\epsilon)p\right)}{2\epsilon},
\label{eq:semi_elasticity}
\end{equation}
with $\epsilon\ll 1$, which rank the parameters by their influence on the stability
margin and on the critical delay. The drag shape parameters enter only the quadratic
term of \eqref{eq:eom} and are absent from the linear operator \eqref{eq:nondim_dde},
so their linear semi-elasticities vanish to numerical precision, which serves as an
internal consistency check on the differencing. The robustness assessment tracks the
ordinal coordinates \eqref{eq:perm_entropy}-\eqref{eq:complexity} of a stable and an
unstable record as the observational noise is increased, verifying that the two
remain separated until the noise overwhelms the signal, and it evaluates the steady
depth response to a small periodic modulation of the carried gas swept in frequency,
which peaks near the slow buoyancy resonance set by \eqref{eq:omega0} and decays
toward ordinary breathing rates, so that the closed loop acts as a low-pass filter on
tidal volume.

\section{Results}

\begin{figure}[H]
\centering
\includegraphics[width=\linewidth]{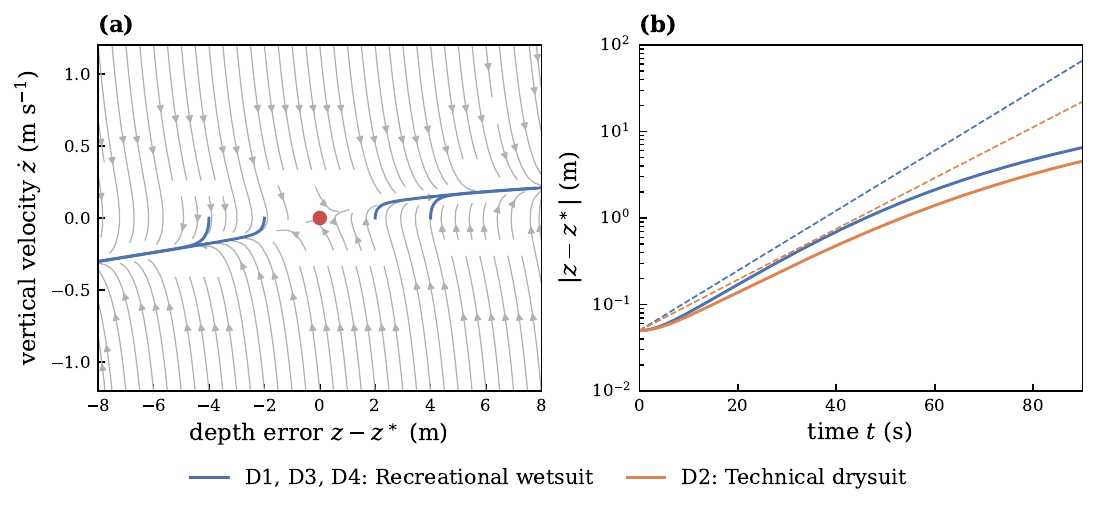}
\caption{Open-loop instability of the uncontrolled buoyancy plant with the carried
gas fixed at the neutral load. (a) Phase portrait in the depth-error and vertical
velocity plane showing the saddle at neutral, with representative trajectories.
(b) Growth of $\lvert z - z^\ast\rvert$ against time for the two distinct plants,
with dashed analytic envelopes $0.05\exp(\lambda_+ t)$; the recreational plant
(D1, D3, D4) and the drysuit plant (D2) are the only two curves.}
\label{fig:openloop}
\end{figure}

With the carried gas held fixed at its neutral load, the uncontrolled plant is a
saddle for every configuration, as shown in Figure~\ref{fig:openloop}. The three
divers that share the recreational hull and weighting at the $15\,$m target,
namely D1, D3, and D4, collapse onto a single open-loop plant with negative
stiffness $\beta = 1.5951\,$N\,m$^{-1}$, buoyancy authority
$\gamma = 4040.58\,$N\,m$^{-3}$, instability rate
$\omega_0 = 0.126297\,$s$^{-1}$, and saddle eigenvalues
$\lambda_+ = +0.079825\,$s$^{-1}$ and $\lambda_- = -0.199825\,$s$^{-1}$,
corresponding to an unstable-manifold e-folding time of $12.5274\,$s and a neutral
gas load $q^\ast = 3.8996\,$L. The technical drysuit D2 at the $30\,$m target forms
the only distinct plant, with $\beta = 1.4971\,$N\,m$^{-1}$,
$\gamma = 2528.27\,$N\,m$^{-3}$, $\omega_0 = 0.111696\,$s$^{-1}$,
$\lambda_+ = +0.067678\,$s$^{-1}$, $\lambda_- = -0.184345\,$s$^{-1}$, an e-folding
time of $14.7759\,$s, and $q^\ast = 11.7317\,$L. Figure~\ref{fig:openloop}(a)
shows the saddle phase portrait, and Figure~\ref{fig:openloop}(b) shows the
divergence of $\lvert z - z^\ast\rvert$ tracking the analytic envelopes
$0.05\exp(\lambda_+ t)$, with D1, D3, and D4 coincident and D2 diverging more
slowly.

\begin{figure}[H]
\centering
\includegraphics[width=\linewidth]{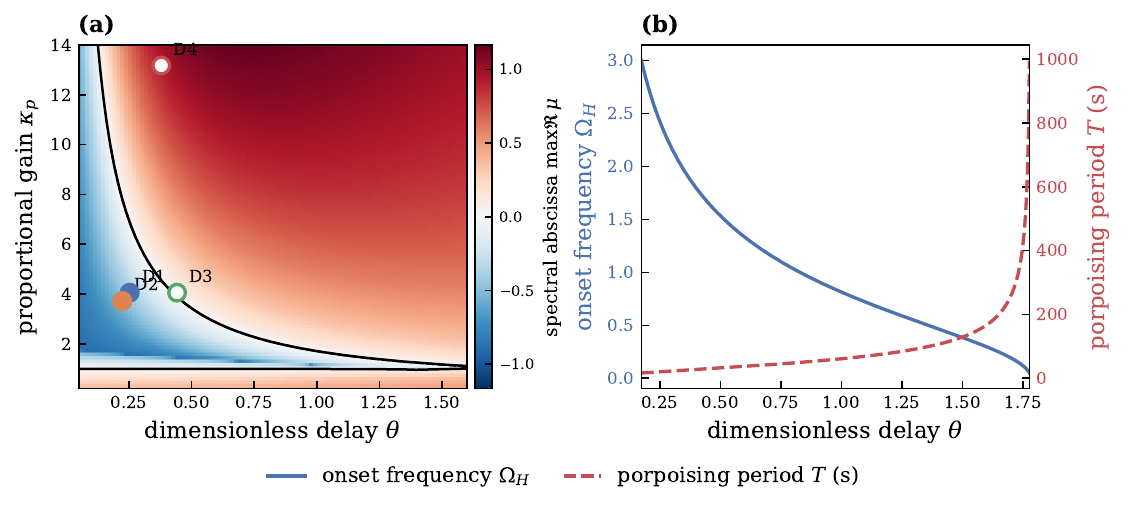}
\caption{Linear stability chart and porpoising onset. (a) Spectral abscissa over
the dimensionless delay $\theta$ and proportional gain $\kappa_p$ at the baseline
slice, with the solid curve the zero level of the pseudospectral abscissa, the
dashed curve the closed-form Hopf boundary, and the four divers marked with their
own $\kappa_d$. (b) Onset frequency $\Omega_H$ and porpoising period $T$ along the
boundary as functions of $\theta$.}
\label{fig:stability}
\end{figure}

The linear stability chart in Figure~\ref{fig:stability} maps the pseudospectral
spectral abscissa over the dimensionless delay and proportional gain plane at the
baseline slice $\zeta = 0.4751$, $\kappa_d = 0.8318$. Along the Hopf boundary the
critical proportional gain falls from $\kappa_{p,\mathrm{crit}} = 10.0944$ at
$\theta = 0.1733$ to $1.1253$ at $\theta = 1.5544$, the crossing frequency falls
from $\Omega_H = 3.0000$ to $0.3383$, and the associated onset period rises from
$16.5831\,$s to $147.0365\,$s, as plotted in Figure~\ref{fig:stability}(b). Using
each diver's own control parameters, the rightmost characteristic root is
$-0.38540 - 1.78874\mathrm{i}$ for D1 and $-0.43884 - 1.63881\mathrm{i}$ for D2,
placing both in the stable hover regime with associated periods of $27.812\,$s and
$34.325\,$s, while the root is $+0.16674 + 1.57507\mathrm{i}$ for D3 and
$+0.98285 - 2.52324\mathrm{i}$ for D4, placing both in the porpoising regime with
associated periods of $31.585\,$s and $19.716\,$s. The four operating points are
marked in Figure~\ref{fig:stability}(a).

\begin{figure}[H]
\centering
\includegraphics[width=\linewidth]{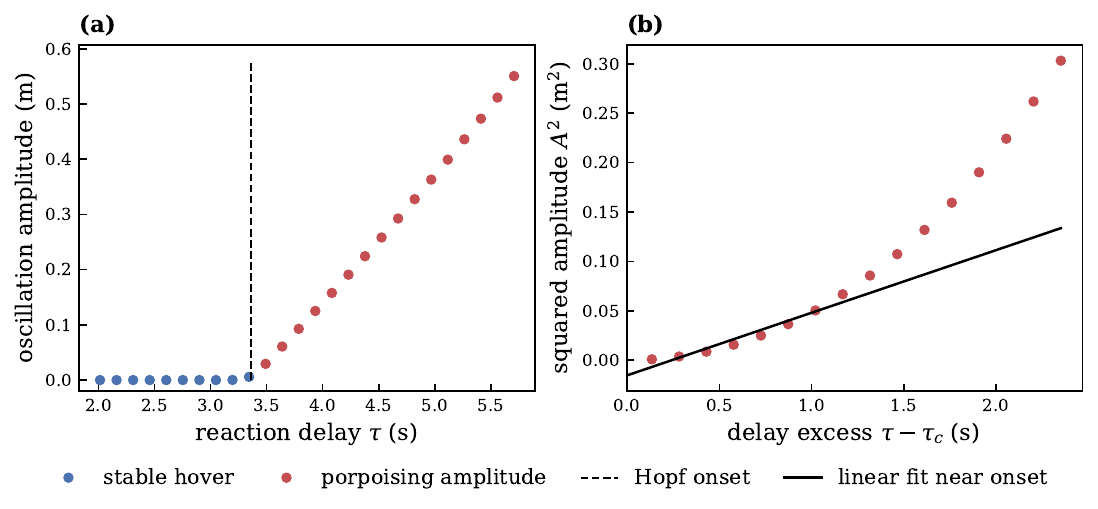}
\caption{Hopf bifurcation of the controlled baseline plant. (a) Steady
oscillation amplitude against reaction delay, with the dashed line marking the
pseudospectral onset $\tau_c = 3.356996\,$s. (b) Squared amplitude against delay
excess $\tau - \tau_c$ near onset with the linear fit.}
\label{fig:bifurcation}
\end{figure}

The bifurcation of the baseline diver is shown in Figure~\ref{fig:bifurcation}. The
spectral abscissa crosses zero at a critical delay $\tau_c = 3.356996\,$s,
equivalently $\theta_c = 0.423979$, with onset frequency $\Omega_H = 1.724779$ and
onset period $28.8438\,$s. The steady peak-to-peak amplitude is zero on the hover
side and grows continuously beyond onset, reaching $0.55064\,$m at
$\tau = 5.7069\,$s in Figure~\ref{fig:bifurcation}(a). A least-squares fit of the
squared amplitude against the delay excess near onset in
Figure~\ref{fig:bifurcation}(b) gives a slope of
$\mathrm{d}A^2/\mathrm{d}\tau = 0.063419\,$m$^2$\,s$^{-1}$ with a coefficient of
determination of $0.94551$.

\begin{figure}[H]
\centering
\includegraphics[width=\linewidth]{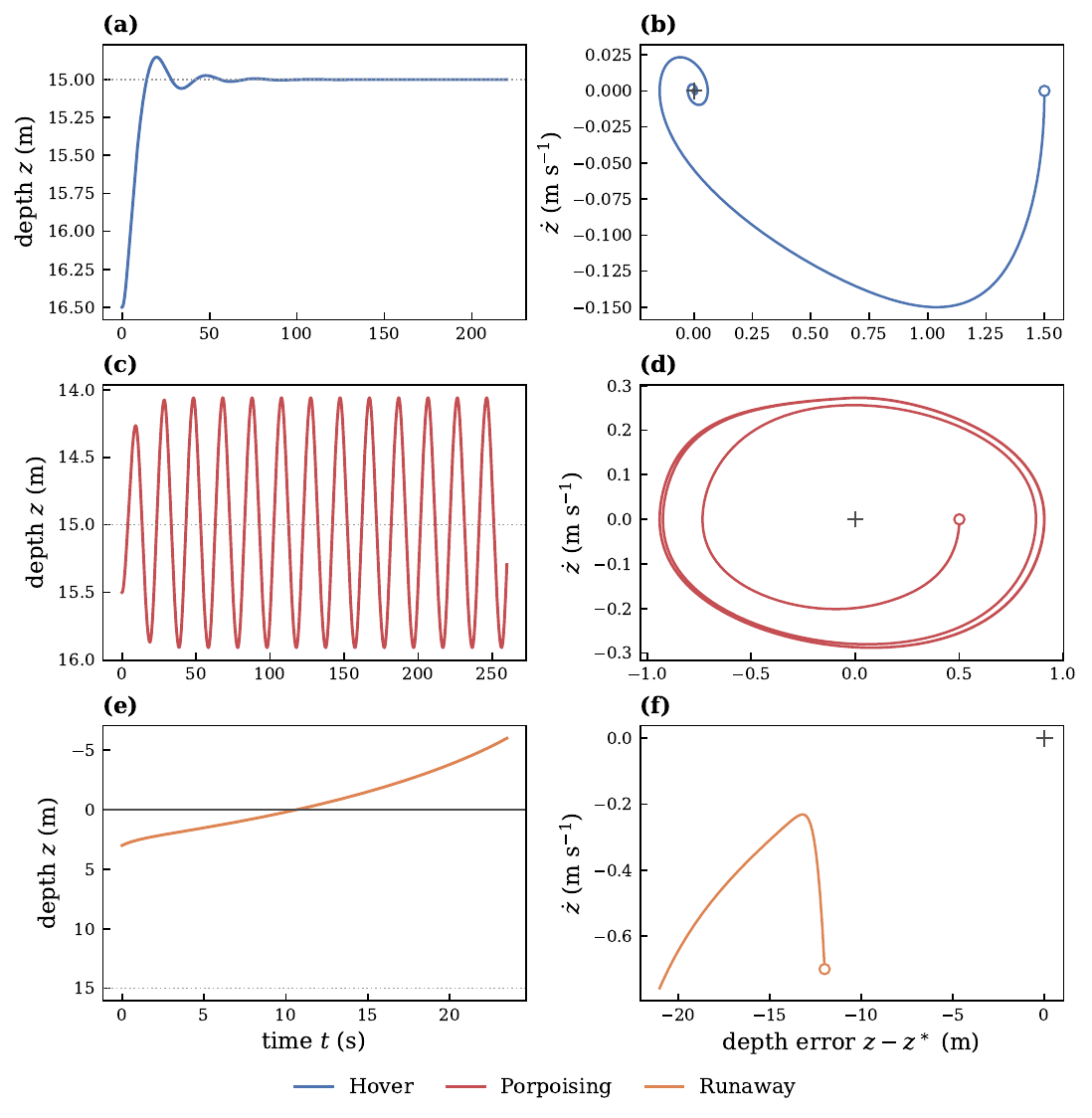}
\caption{Closed-loop regimes. (a, b) Hover for D1, (c, d) porpoising for D4, and
(e, f) runaway for D1, each shown as a depth history and a phase portrait in the
depth-error and vertical velocity plane. The dotted lines mark the neutral depth;
crosses mark the neutral state.}
\label{fig:regime}
\end{figure}

Figure~\ref{fig:regime} collects one representative trajectory for each closed-loop
regime. Started $1.50\,$m from neutral, the hover case (D1) settles to a final
depth of $15.000\,$m with a late peak-to-peak amplitude of $0.001\,$m over a
$220\,$s horizon, visiting depths in $[14.85, 16.50]\,$m. Started $0.50\,$m from
neutral, the porpoising case (D4) sustains a limit cycle with period $19.805\,$s
and late peak-to-peak amplitude $1.849\,$m over a $260\,$s horizon, visiting depths
in $[14.06, 15.91]\,$m. Started $12.00\,$m shallow of neutral with an upward
velocity of $0.70\,$m\,s$^{-1}$, the runaway case (D1) reaches the surface within
$90\,$s, the depth spanning $[-6.01, 3.00]\,$m and the run ending at $-6.011\,$m
with both the surface-reaching and neutral-escape conditions met. The depth
histories and phase portraits appear in Figure~\ref{fig:regime}(a)--(f).

\begin{figure}[H]
\centering
\includegraphics[width=\linewidth]{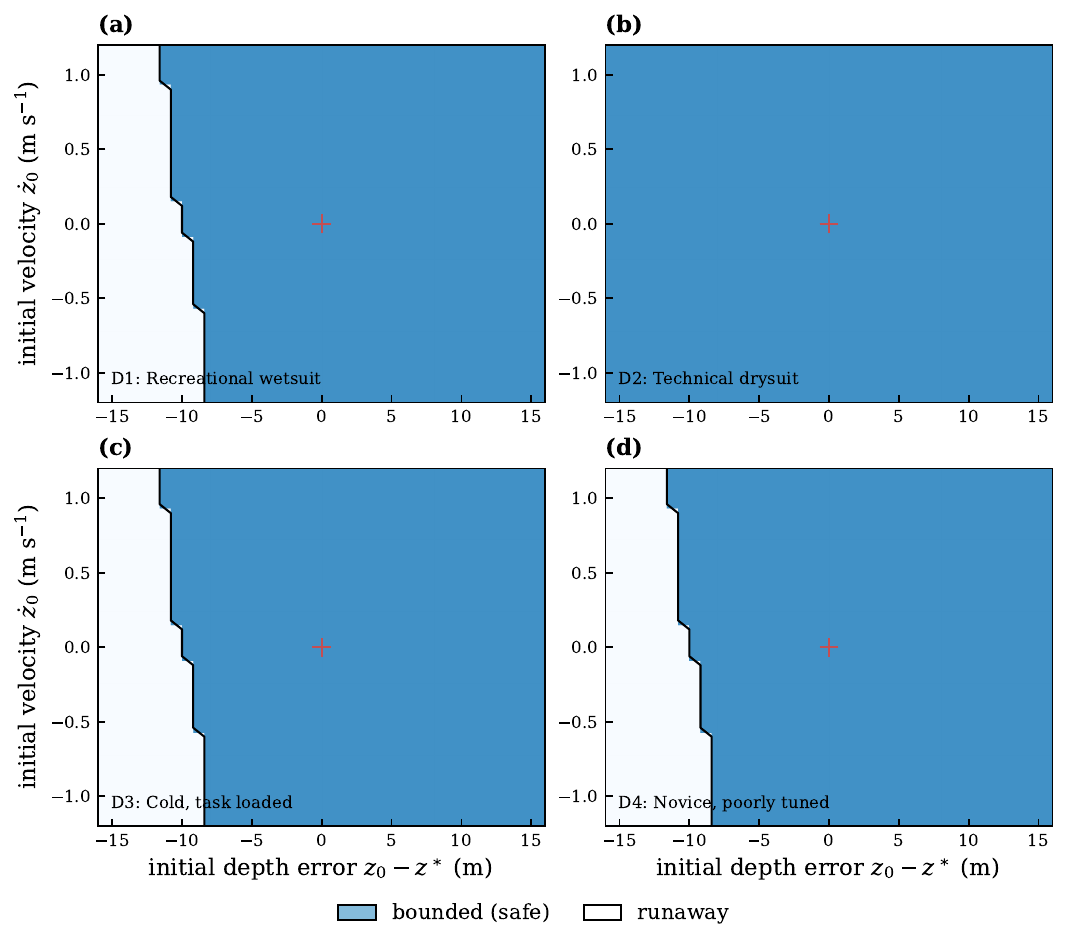}
\caption{Safe operating envelope of the buoyancy loop. Bounded (shaded) and
runaway (white) initial conditions in the plane of initial depth error and initial
vertical velocity for (a) D1, (b) D2, (c) D3, and (d) D4. The cross marks the
neutral state.}
\label{fig:envelope}
\end{figure}

The safe operating envelope in Figure~\ref{fig:envelope} classifies a
$41 \times 41$ grid of initial depth error in $[-16, 16]\,$m and vertical velocity
over a $150\,$s horizon as bounded or runaway. The bounded fraction of the sampled
plane is $0.8007$ for D1, D3, and D4, and $1.0000$ for D2. At zero initial
velocity, the largest recoverable depth error on the deep side is $+16.000\,$m for
all four divers, spanning the full sampled range, while on the shallow side it is
$-9.600\,$m for D1, D3, and D4 and $-16.000\,$m for D2. The runaway region for the
recreational configurations occupies a wedge at shallow, ascending starts, absent
for D2, as seen across Figure~\ref{fig:envelope}(a)--(d).

\begin{figure}[H]
\centering
\includegraphics[width=\linewidth]{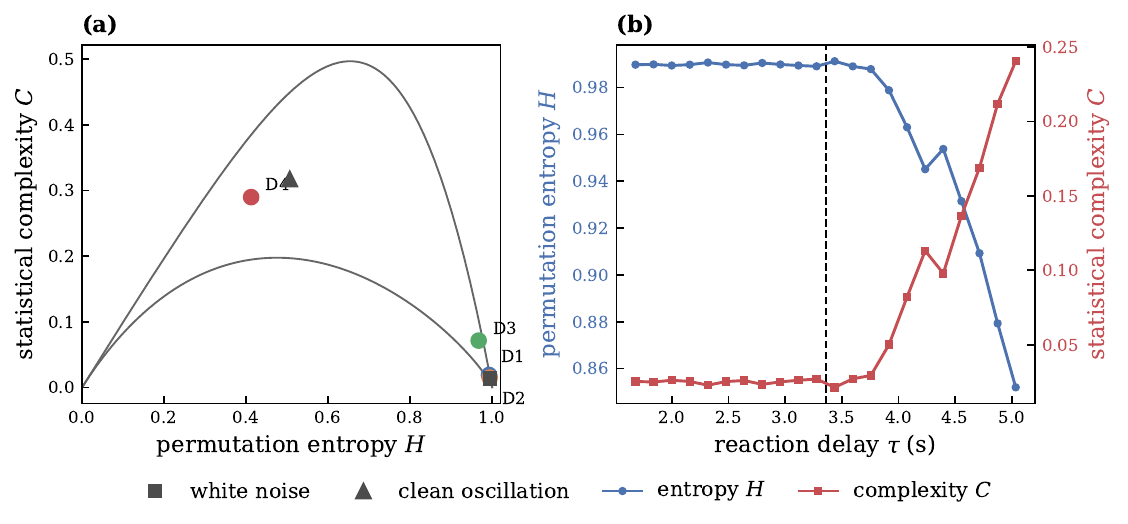}
\caption{Complexity-entropy analysis of the buoyancy regimes. (a) Normalized
permutation entropy $H$ and statistical complexity $C$ for the four divers, a
white-noise reference, and a clean oscillation, with the extremal complexity curves
bounding the plane. (b) $H$ and $C$ along a delay ramp through onset; the dashed
line marks $\tau_c$.}
\label{fig:complexity}
\end{figure}

The complexity-entropy analysis in Figure~\ref{fig:complexity} places the four
divers, a white-noise reference, and a clean-oscillation reference on the plane of
normalized permutation entropy $H$ and statistical complexity $C$ at embedding
dimension six and unit lag with $0.020\,$m observational noise. The coordinates are
$(H, C) = (0.99225, 0.01881)$ for D1, $(0.99383, 0.01489)$ for D2,
$(0.96691, 0.07121)$ for D3, and $(0.41212, 0.28970)$ for D4, with the white-noise
reference at $(0.99436, 0.01369)$ and the clean oscillation at
$(0.50622, 0.31723)$, as shown in Figure~\ref{fig:complexity}(a). Along a delay
ramp through onset, Figure~\ref{fig:complexity}(b), the entropy holds near $0.990$
and the complexity near $0.025$ for delays up to about $3.76\,$s, then the entropy
falls to $0.85202$ and the complexity rises to $0.24061$ at $\tau = 5.0355\,$s.

\begin{figure}[H]
\centering
\includegraphics[width=\linewidth]{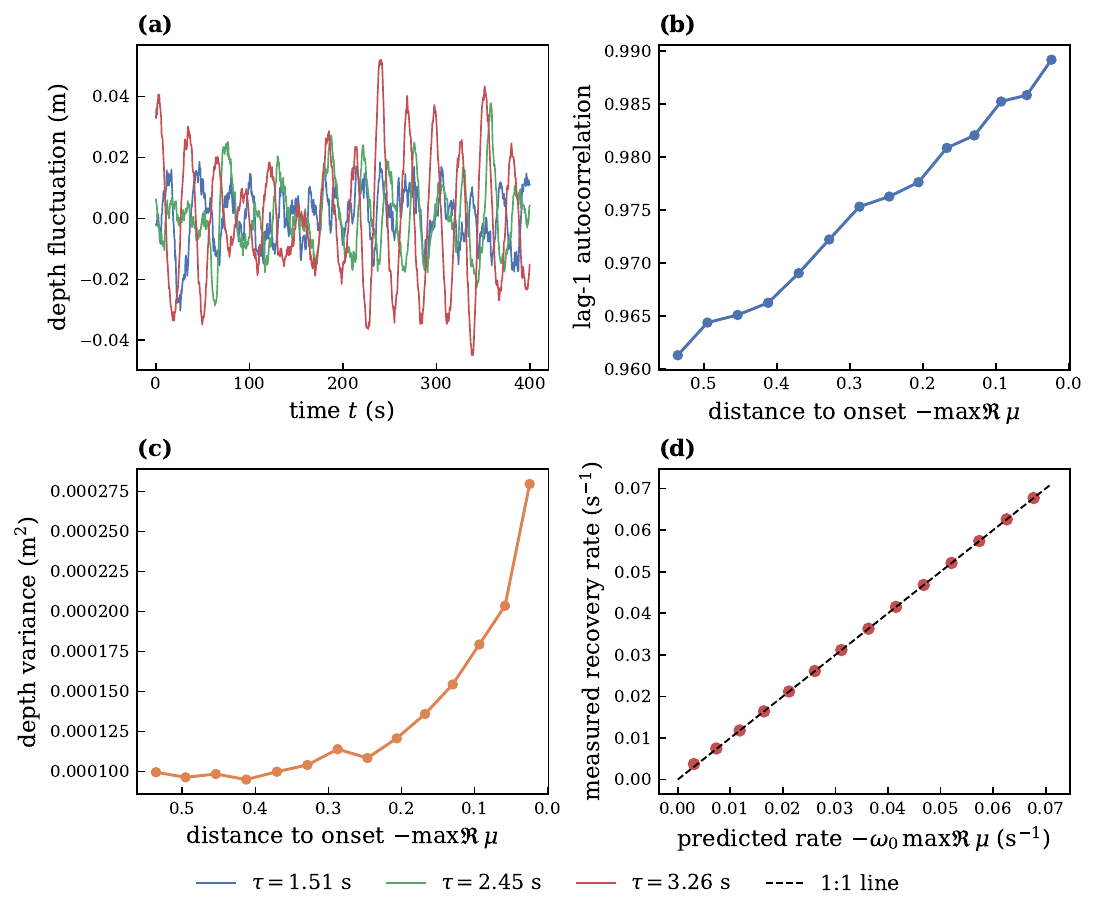}
\caption{Critical slowing down as early warning of onset. (a) Depth-fluctuation
records at three delays. (b) Lag-one autocorrelation, (c) depth variance, and
(d) measured recovery rate against the pseudospectral prediction, each versus the
distance to onset $-\max\operatorname{Re}\mu$; the dashed line in (d) is the
identity.}
\label{fig:earlywarning}
\end{figure}

The critical-slowing-down indicators in Figure~\ref{fig:earlywarning} are computed
on the stable side up to $0.97\tau_c$. As the delay increases from $1.5106\,$s to
$3.2563\,$s the spectral abscissa rises from $-0.53609$ to $-0.02436$, the lag-one
autocorrelation rises from $0.9613$ to $0.9892$, the depth variance rises from
$9.958\times 10^{-5}\,$m$^2$ to $2.797\times 10^{-4}\,$m$^2$, and the fitted
recovery rate falls from $0.06771\,$s$^{-1}$ to $0.00373\,$s$^{-1}$. The Kendall
rank correlation against decreasing distance to onset is $+1.0000$
($p = 2.29\times 10^{-11}$) for the autocorrelation, $+0.8681$
($p = 5.53\times 10^{-7}$) for the variance, and $-1.0000$
($p = 2.29\times 10^{-11}$) for the recovery rate. Figure~\ref{fig:earlywarning}(d)
shows the fitted recovery rate lying on the identity line against the
pseudospectral prediction $-\omega_0\max\operatorname{Re}\mu$.

\begin{figure}[H]
\centering
\includegraphics[width=\linewidth]{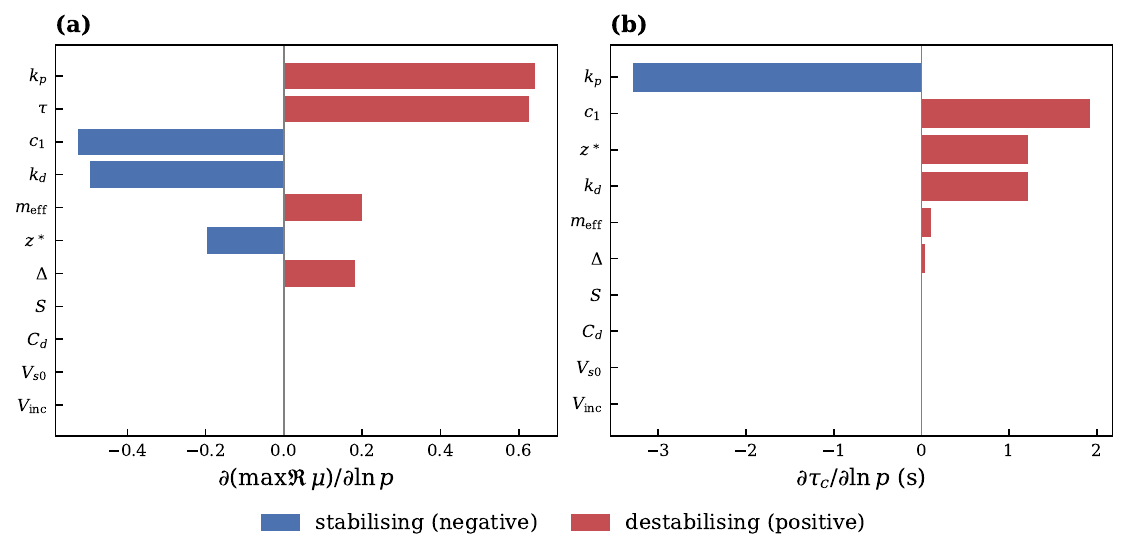}
\caption{Parameter sensitivity of stability and onset for the baseline diver.
(a) Semi-elasticity of the spectral abscissa and (b) semi-elasticity of the
critical delay to each physical parameter, ranked, with stabilizing entries in one
color and destabilizing entries in the other.}
\label{fig:sensitivity}
\end{figure}

The parameter sensitivities in Figure~\ref{fig:sensitivity} are central-difference
semi-elasticities about the baseline, for which the spectral abscissa is $-0.385403$
and the critical delay is $3.356996\,$s. The abscissa semi-elasticities in
Figure~\ref{fig:sensitivity}(a) are $+0.6401$ for $k_p$, $+0.6244$ for $\tau$,
$-0.5249$ for $c_1$, $-0.4954$ for $k_d$, $+0.1979$ for $m_\mathrm{eff}$, $-0.1955$
for $z^\ast$, and $+0.1822$ for $\Delta$, with $C_d$, $S$, $V_\mathrm{inc}$, and
$V_{s0}$ identically zero. The critical-delay semi-elasticities in
Figure~\ref{fig:sensitivity}(b) are $-3.2823\,$s for $k_p$, $+1.9182\,$s for $c_1$,
$+1.2147\,$s for $z^\ast$, $+1.2137\,$s for $k_d$, $+0.1125\,$s for
$m_\mathrm{eff}$, and $+0.0378\,$s for $\Delta$, again with $C_d$, $S$,
$V_\mathrm{inc}$, and $V_{s0}$ identically zero.

\begin{figure}[H]
\centering
\includegraphics[width=\linewidth]{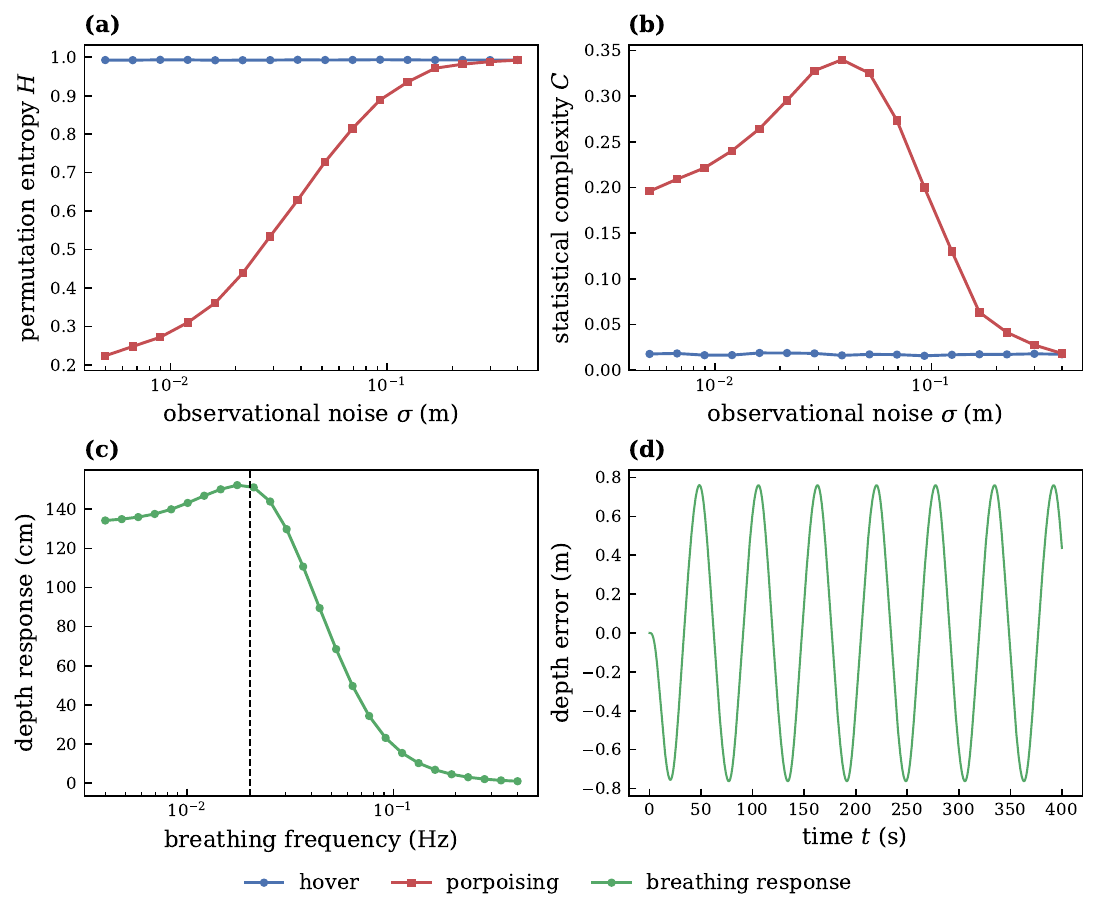}
\caption{Robustness to observational noise and breathing forcing. (a) Permutation
entropy and (b) statistical complexity of hover and porpoising records against
observational noise. (c) Steady depth response to a breathing modulation of the
carried gas swept in frequency, with the dashed line at the undamped corner, and
(d) a representative depth-error record under breathing forcing.}
\label{fig:robustness}
\end{figure}

The robustness analysis in Figure~\ref{fig:robustness} tracks ordinal separation
under observational noise and the response to breathing modulation. At the smallest
noise level $\sigma = 0.0050\,$m the entropy is $0.9927$ for hover and $0.2238$ for
porpoising and the complexity is $0.0178$ and $0.1960$, a separation of
$\lvert\Delta H\rvert = 0.7689$; as $\sigma$ increases to $0.4000\,$m the
porpoising entropy rises to $0.9924$ and the separation falls to
$\lvert\Delta H\rvert = 0.0005$, while the porpoising complexity peaks at $0.3399$
near $\sigma = 0.0386\,$m before declining, and the hover coordinates hold near
$H = 0.993$ and $C = 0.017$ throughout, as shown in
Figure~\ref{fig:robustness}(a) and (b). For a $0.80\,$L breathing modulation of the
carried gas, Figure~\ref{fig:robustness}(c) gives an undamped corner of
$0.0201\,$Hz, a damped resonance of $0.0149\,$Hz, a swept-response peak at
$0.0175\,$Hz with a peak depth response of $152.3046\,$cm, and a response of
$2.6105\,$cm at $0.25\,$Hz.

\section{Discussion}

The mechanism is a delayed feedback stabilization of an inherently unstable
equilibrium, and every result follows from that single fact. With the carried gas
frozen at the neutral load the plant is a saddle whose two eigenvalues are real, one
positive and one negative, so the uncontrolled buoyancy dynamics contain no
oscillation of any kind; the instability rate $\omega_0 = 0.126\,$s$^{-1}$ sets an
e-folding time near twelve seconds for the recreational plant and near fifteen
seconds for the drysuit, over which an unattended displacement grows monotonically.
The oscillation that defines porpoising is therefore not a property of the plant but
is generated entirely by the retarded control: the rightmost roots of the closed
loop carry large imaginary parts, $\pm 1.575$ for the cold, task-loaded diver and
$\pm 2.523$ for the poorly tuned novice in dimensionless units, that are absent from
the open-loop spectrum. This places diver porpoising in the same class as the
delayed stabilization of the inverted pendulum and of quiet human stance, in which a
finite sensorimotor latency turns an otherwise controllable unstable equilibrium into
a self-sustained oscillation \cite{Stepan2009,Milton2009,Masani2006}.

The stability chart exposes the two-boundary structure characteristic of this class.
The stable tongue is bounded below by a static limit at unit proportional gain, where
the delayed proportional feedback is just strong enough to overcome the negative
buoyancy stiffness, and above by the Hopf curve, where the delay converts excess gain
into oscillation. Along the Hopf boundary the crossing frequency falls from
$\Omega_H = 3.00$ at short delay to $0.34$ at long delay while the onset period rises
from $16.6\,$s to $147.0\,$s, so a slower reaction produces slower and larger
porpoising rather than faster, a counterintuitive but generic feature of
delay-induced Hopf bifurcations in infinite-dimensional systems
\cite{Diekmann1995}. The two unstable divers reach the boundary by distinct routes
that the sensitivity ranking makes precise: the novice sits deep in the unstable
region through an excessive proportional gain, the largest single destabilizing
influence with an abscissa semi-elasticity of $+0.64$ and a critical-delay
semi-elasticity of $-3.28\,$s, whereas the cold, task-loaded diver crosses through a
deficient derivative gain, the second most stabilizing parameter at $-0.50$, which
represents the sluggish rate sensing of a cold and distracted operator. That the drag
shape parameters and the volume terms carry identically zero linear sensitivity is
not a numerical accident but a consequence of the reduction to four dimensionless
groups: only quantities entering those groups can move the spectrum, and the vanishing
entries confirm the reduction rather than reporting a physical effect.

Past onset the bifurcation is supercritical, and its amplitude law is diagnostic of
the saturating physics. The branch emerges continuously from zero at
$\tau_c = 3.357\,$s, and across the entire computed range the steady amplitude is
very nearly proportional to the delay excess, with the ratio $A/(\tau-\tau_c)$ holding
near $0.22$ from the first resolved point outward and drifting upward by only about
seven percent at the largest excess. This linear growth, rather than the square-root
growth of the generic Hopf normal form \cite{Guckenheimer1983}, is the signature of
the quadratic hydrodynamic drag: a degree-two homogeneous dissipation
$\propto\lvert\dot z\rvert\dot z$ balances the linear energy input of the delayed
feedback at an amplitude that scales linearly with the distance to onset rather than
as its square root. The linear fit of squared amplitude against delay excess, whose
positive slope of $0.063\,$m$^2$\,s$^{-1}$ and coefficient of determination of
$0.945$ we report as evidence of supercriticality, is best read in this light: the
systematic bowing of the data above that straight line is precisely the fingerprint
of the linear-in-excess amplitude law, and a cleaner exponent would require finer
sampling immediately above onset. The nonlinear limit-cycle period of the novice,
$19.8\,$s, sits within one percent of the linear Hopf period of $19.7\,$s, so the
oscillation frequency is set by the delayed feedback and barely shifts with amplitude,
as expected for a soft onset.

Local stability does not, however, determine safety. The safe-envelope analysis
shows a bounded fraction of $0.80$ that is identical for the stable recreational diver
and the porpoising novice, because escape is governed by compensator saturation
rather than by the sign of the rightmost eigenvalue: near neutral both the stable node
and the stable limit cycle remain confined, and runaway occupies a wedge of shallow,
ascending starts where the compensator vents to empty and the loop loses authority.
The recoverable range is correspondingly asymmetric, extending to the full sampled
depth on the deep side but only to $9.6\,$m on the shallow side, and the dangerous
direction is the uncontrolled ascent, in which a diver too shallow and rising exhausts
the compensator and cannot arrest the climb. The drysuit configuration recovers from
every sampled start, a global margin attributable to its larger neutral gas reserve
and deeper target rather than to any difference in linear stability. The distinction
between local decay and global escape, and its control by actuator saturation, is the
practically decisive feature of the model and the one least visible in a purely linear
treatment.

The transition is heralded from the stable side by critical slowing down, and the
model allows the precursor to be tied to the exact spectrum rather than treated as a
heuristic. As the delay approaches its critical value the fitted recovery rate falls
from $0.068\,$s$^{-1}$ to $0.004\,$s$^{-1}$ and lies on the identity line against the
pseudospectral prediction $-\omega_0\max\operatorname{Re}\mu$, so the recovery rate is
a direct estimate of the leading eigenvalue's real part, and its rank correlation with
distance to onset is a perfect $-1.00$. The lag-one autocorrelation rises equally
monotonically with a rank correlation of $+1.00$, while the variance rises with a
weaker correlation of $+0.87$, reproducing the known ordering in which
autocorrelation-based indicators outperform variance as early-warning signals
\cite{Scheffer2009,Dakos2012}. That the recovery rate coincides with the spectral
abscissa gives this canonical early-warning framework an exact reference here, which
is rarely available in the empirical settings where it is usually applied.

The information-theoretic diagnostics describe the same transition in a
model-independent language and reveal a complementary limitation. On the
complexity-entropy plane the two hovering divers sit at the stochastic corner beside
the white-noise reference, because their millimeter-scale residual motion lies far
below the two-centimeter observational noise, so their coordinates report the noise
rather than the dynamics. The porpoising novice sits deep in the ordered region at
entropy $0.41$ and complexity $0.29$, at even lower entropy than the clean-oscillation
reference, which indicates that the strongly anharmonic porpoising cycle populates
fewer ordinal patterns than a pure sinusoid and is in that sense more predictable
\cite{BandtPompe2002,Rosso2007}. Along the delay ramp the coordinate does not move
until the delay exceeds roughly $3.76\,$s, past the linear onset at $3.357\,$s,
because the ordinal estimator responds only once the limit-cycle amplitude clears the
noise floor. The spectral and variance precursors therefore lead the transition from
the stable side while the ordinal indicators lag it and confirm the developed
nonlinear state, and the two families of diagnostic are best understood as
complementary rather than redundant. The robustness analysis bounds the reach of the
ordinal measures: the entropy separation between hover and porpoising collapses from
$0.77$ to below $0.001$ as observational noise floods the porpoising signal, and the
statistical complexity of that signal peaks at an intermediate noise near
$0.04\,$m, so discrimination requires measurement noise well below the meter-scale
porpoising amplitude, a condition that centimeter-resolution depth sensors satisfy by
two orders of magnitude.

Breathing enters the loop as an external forcing, and the response is a clean
low-pass filter distinct from the delay-induced Hopf mechanism. The undamped corner
sits at $0.020\,$Hz and the damped resonance at $0.015\,$Hz, consistent with the
baseline damping ratio through $f_r = f_n\sqrt{1-2\zeta^2}$, more than a decade below
ordinary respiration. A tidal gas exchange near a quarter of a hertz produces only
$2.6\,$cm of depth excursion, whereas the same exchange applied near the resonance
produces more than a meter. Routine breathing therefore couples negligibly to depth,
but slow breathing or task cadences that approach the buoyancy corner can drive
meter-scale excursions through resonance rather than instability, a pathway that
operates even for a diver whose control loop is comfortably stable.

The scope of these conclusions is set by the deliberately reduced character of the
model, which is a toy model in the precise sense that it retains one mechanism and
discards everything not essential to it. The reduction is threefold. Physically, the
model resolves only the vertical degree of freedom and represents the diver as a
point mass in heave, so attitude, pitch, lateral translation, and the coupling
between them are absent, and the trim torque implied by the offset lines of action of
weight and buoyancy is not modeled; the added mass and the water temperature are held
constant, the gas compression is taken as isothermal, and the seawater density is
uniform, each of which is defensible over a limited depth excursion but not across a
full descent. Control-theoretically, the human operator is collapsed to a fixed
linear proportional-derivative law acting through a single constant delay, whereas a
real diver adapts gains with experience, carries a latency that varies with workload
and arousal, senses depth and rate through noisy and quantized channels, and very
likely acts predictively rather than reactively; the saturation of the compensator is
retained because it governs escape, but the finite inflation and venting rates of a
real buoyancy device, the discrete breath-by-breath nature of lung-volume control,
and any dead band in perception are not. Parametrically, the four canonical divers are
constructed to span stable and unstable behavior rather than measured against
instrumented dives, so their numerical values are illustrative and the quantitative
thresholds, the critical delay near three and a third seconds, the onset period near
twenty-nine seconds, and the meter-scale porpoising amplitude, should be read as the
behavior of the model at representative settings rather than as predictions for any
individual. Numerically, the time integration is globally second order because the
constant initial history injects a derivative discontinuity that propagates through
the delay intervals, which is adequate for the reported invariants but too coarse to
resolve the amplitude exponent immediately above onset, and the ordinal and
early-warning diagnostics inherit the sampling interval and noise level chosen here.
None of these simplifications is incidental; each is a modeling decision that trades
fidelity for the isolation of a single pathway, and each also marks a direction in
which the model can be extended without altering its core. Pitch and lateral degrees
of freedom, a state-dependent or stochastic delay, an adaptive or predictive
controller, rate-limited actuation, and non-isothermal gas dynamics are the natural
next steps, and the diagnostic apparatus assembled here, the pseudospectral spectrum,
the amplitude branch, the safe envelope, the critical-slowing-down indicators, and the
complexity-entropy coordinates, transfers to each of them without modification.
Within its stated limits the model is not a description of diving practice but a
minimal dynamical account of why delayed regulation of an inherently unstable buoyancy
equilibrium should produce hovering, porpoising, and runaway at all, and of the form
each takes.

\section{Conclusion}

This work reframed a familiar diving hazard as a problem in delayed nonlinear
control and showed that a single reduced model reproduces its full behavioral
repertoire. Treating the diver as a delayed proportional-derivative regulator of an
inherently unstable buoyancy equilibrium, we derived the governing delay differential
equation from first principles, located its stability boundary by pseudospectral
collocation of the generator, and traced the nonlinear branch and its basin by
delay-aligned time integration, complementing these with information-theoretic and
critical-slowing-down diagnostics. The value of the account is that it renders
porpoising, hovering, and runaway as outcomes of one mechanism whose control
parameters are identifiable and whose onset is an exactly characterized supercritical
delay-induced Hopf bifurcation, and that it equips the study with a transferable set of diagnostics tying statistical precursors to the leading eigenvalue. The framework is deliberately minimal, and its extension to attitude, adaptive and stochastic delay, rate-limited actuation, and non-isothermal gas dynamics, together with validation against instrumented dives, defines the path from this idealized model toward a quantitative description of human buoyancy control.
\section*{Acknowledgements}
The authors used Claude Sonnet~5 (Anthropic, PBC) solely as a
writing-assistance tool to refine English vocabulary and grammar
during the preparation of this manuscript. All scientific content,
interpretations, analyses, conclusions, and any remaining linguistic
imperfections are the sole responsibility of the authors. We further
thank the Indonesian Subaquatic Sport Association (POSSI) instructors
Jemi Godjali (INA.F00.B3.0091), Saep Saepudin (INA.F00.B1.0396), and
Subagiyo (INA.F00.B1.0412) for their guidance on scuba diving
practice.

\section*{Funding}
This study was funded by the ITB 3P Research Program 2026 (Talenta
Unggul Scheme) through the Directorate of Research and Innovation,
Bandung Institute of Technology (Project ID: DRI.PN-6-64-2026) to
S.H.S.H. and A.P.H.

\section*{Author Contributions}
\textbf{S.H.S.H.}: Conceptualization, Methodology, Software, Validation, Formal analysis, Investigation, Data curation, Visualization, Project administration, Funding acquisition, Writing -- original draft. \textbf{F.A.R.A.}: Supervision, Writing -- review \& editing. \textbf{I.P.A.}: Resources, Supervision, Writing -- review \& editing. \textbf{F.K.}: Supervision, Writing -- review \& editing. \textbf{A.P.H.}: Supervision, Funding acquisition, Writing -- review \& editing. \textbf{K.A.S.}: Supervision, Funding acquisition, Writing -- review \& editing. \textbf{R.S.}: Supervision, Writing -- review \& editing. \textbf{D.E.I.}: Supervision, Writing -- review \& editing. All authors read and approved the final manuscript.

\section*{Data Availability}
The Python scripts are available at \url{https://github.com/sandyherho/suppl_scuba_bouyancy}, and the figures and computational notes are archived at \url{https://doi.org/10.17605/OSF.IO/VNMWS}, both under the MIT license. All results are fully reproducible using the provided code.

\end{document}